\documentclass[aps,prb,showpacs,twocolumn,floats]{revtex4}
\usepackage{graphicx}
\usepackage{subfigure}
\usepackage{epsfig}
\usepackage{dcolumn}
\usepackage{bm}
\usepackage[ansinew]{inputenc}
\usepackage{amsmath}
\usepackage{amsthm}
\usepackage[T1]{fontenc}
\usepackage{amssymb}
\usepackage{amsfonts}
\usepackage[english]{babel}
\usepackage{enumitem}

\begin{document}

\title{Enhanced spin tunneling in a molecular magnet mixed with a superconductor}
\date{\today}
\author{J. Tejada, R. Zarzuela, A. Garc\'{i}a-Santiago}
\affiliation{Grup de Magnetisme, Departament de F\'{i}sica Fonamental, Facultat de F\'{i}sica, Universitat de Barcelona, Mart\'{i} i Franqu\`{e}s 1, 08028 Barcelona, Spain}
\author{ I. Imaz$^1$,  J. Espin$^1$, D. Maspoch$^{1,2}$}
\affiliation{$^1$Institut Catal\`{a} de Nanotecnologia, ICN2, Esfera Universitat Aut\'{o}noma Barcelona (UAB), Campus UAB, 08193 Bellaterra,  Spain\\$^2$Instituci\'{o} Catalana de Recerca i Estudis Avan\c{c}ats (ICREA), 08100 Barcelona, Spain}
\author{E. M. Chudnovsky}
\affiliation{Physics Department, Lehman College,
The City University of New York, 250 Bedford Park Boulevard West, Bronx, NY 10468-1589, USA}

\begin{abstract}
We report characterization and magnetic studies of mixtures of micrometer-size ribbons of Mn$_{12}$ acetate and micrometer-size particles of YBaCuO superconductor. Extremely narrow zero-field spin-tunneling resonance has been observed in the mixtures, pointing to the absence of the inhomogeneous dipolar broadening. It is attributed to the screening of the internal magnetic fields in the magnetic particles by Josephson currents between superconducting grains surrounding the particles.
\end{abstract}


\maketitle

\section{Introduction}\label{introduction}
Molecular magnets have been at the forefront of research on quantum spin phenomena at the nanoscale \cite{Springer}. They provide an ultimate limit of the miniaturization of magnetic memory and are promising candidates for qubits -- elements of quantum computers \cite{Nanotech-2001}. Quantum effects observed in molecular magnets include quantum tunneling of the magnetic moment \cite{Friedman-PRL1996,Hernandez-EPL1996,Barbara-Nature1996,MQT-book}, topological Berry phase \cite{WS}, quantum magnetic deflagration \cite{CCNY-PRL2005,UB-PRL2005}, and Rabi oscillations \cite{Schledel-PRL2008,Bertaina-Nature2008}. Most recently it was demonstrated that measurement of the electric current through a magnetic molecule permits readout of quantum states of an individual atomic nucleus \cite{Wern-NatureNano2013,Wern-ASC-Nano2013}. 

In molecular magnets the inversed population of spin energy levels can be created by simply applying the magnetic field. It was suggested some time ago that this can be used to achieve superradiance and laser effect in molecular magnets \cite{chugar-SR,tejada-SR}. Besides fundamental interest, such effects would be of significant practical interest as they would provide sources of coherent electromagnetic radiation in the frequency range of a few hundred GHz which are difficult to obtain by other methods. The subsequent studies \cite{FriedmanSR-PRB2004}, however, revealed that inhomogeneous broadening of spin levels in molecular magnets is a great impediment on the way of achieving coherent radiation. It originates from dipolar and hyperfine interactions, as well as from $D$-strains and $g$-strains \cite{Park-PRB2001}. Typically observed widths of the resonances are in the ball park of $1$kOe. The zero-field resonance stands out because it is not subject to $D$-strains and $g$-strains and because it also shows the absence of hyperfine broadening \cite{Friedman-PRB1998}, apparently due to the fast transitions between nuclear states on a time-scale of a typical field-sweep experiment. For conventional Mn$_{12}$ acetate, the typical width of the zero-field resonance is in the ball park of $300$ Oe. 

In this Letter we report experimental attempt to reduce the width of the zero-field spin-tunneling resonance by mixing micron-size ribbons of Mn$_{12}$ acetate with micron-size grains of YBaCuO. Mn$_{12}$ ribbons with triclinic short-range crystal structure (see below) exhibit more narrow zero-field resonance than conventional Mn$_{12}$ acetate. Further reduction of the width of the resonance requires elimination of the dipolar broadening. Here we explore the idea of screening the internal dipolar magnetic fields in Mn$_{12}$ particles by superconducting currents in the YBaCuO grains surrounding the particles \cite{EC-Friedman-PRL2000}. In accordance with our expectation, we observed a pronounced magnetic relaxation in the vicinity of zero field and the width of the zero-field resonance reduced to values as low as $25$ Oe. 

\begin{figure}[htbp!]
\includegraphics[width=8.0cm,angle=0]{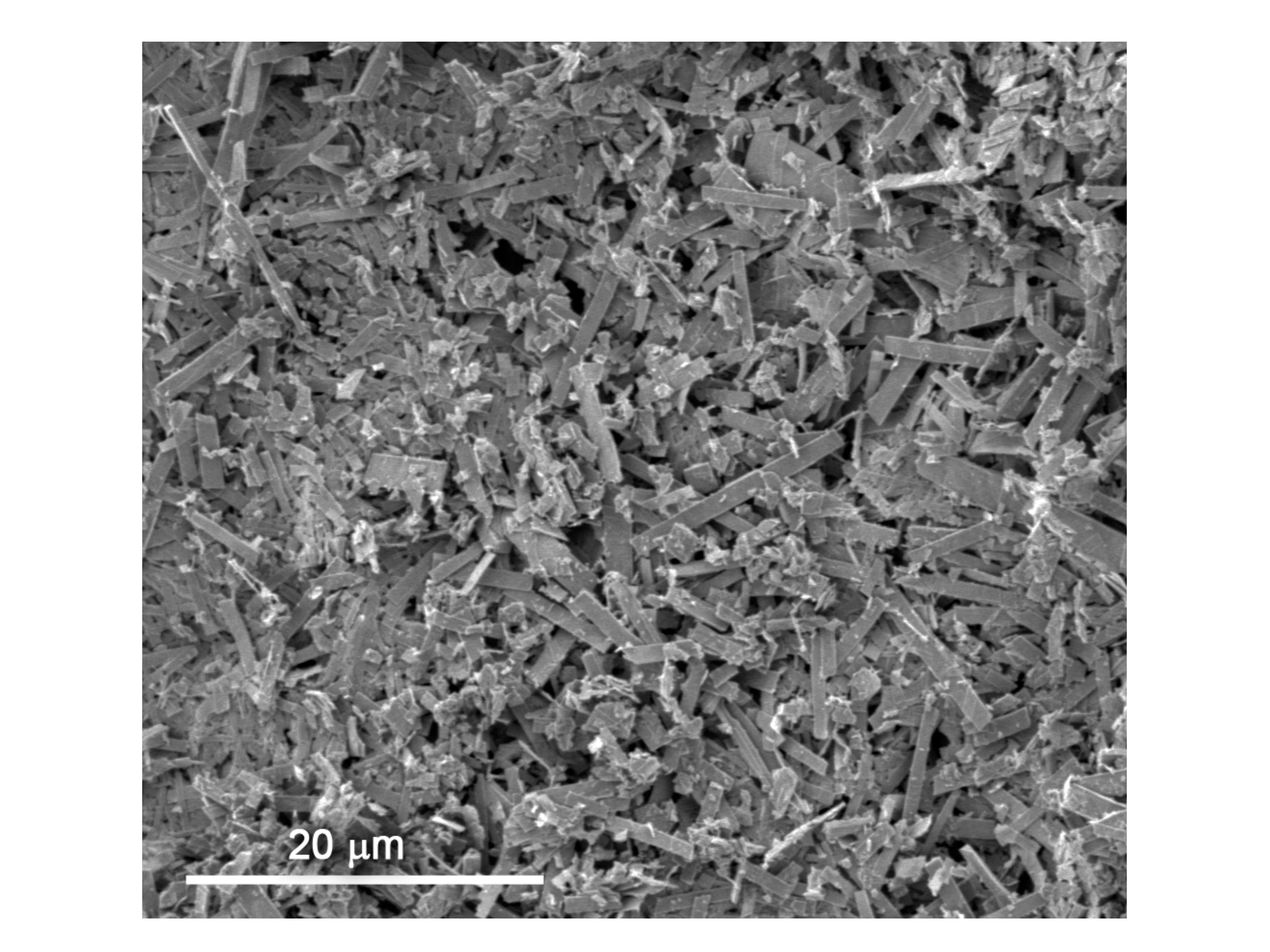}{a}
\includegraphics[width=7.3cm,angle=0]{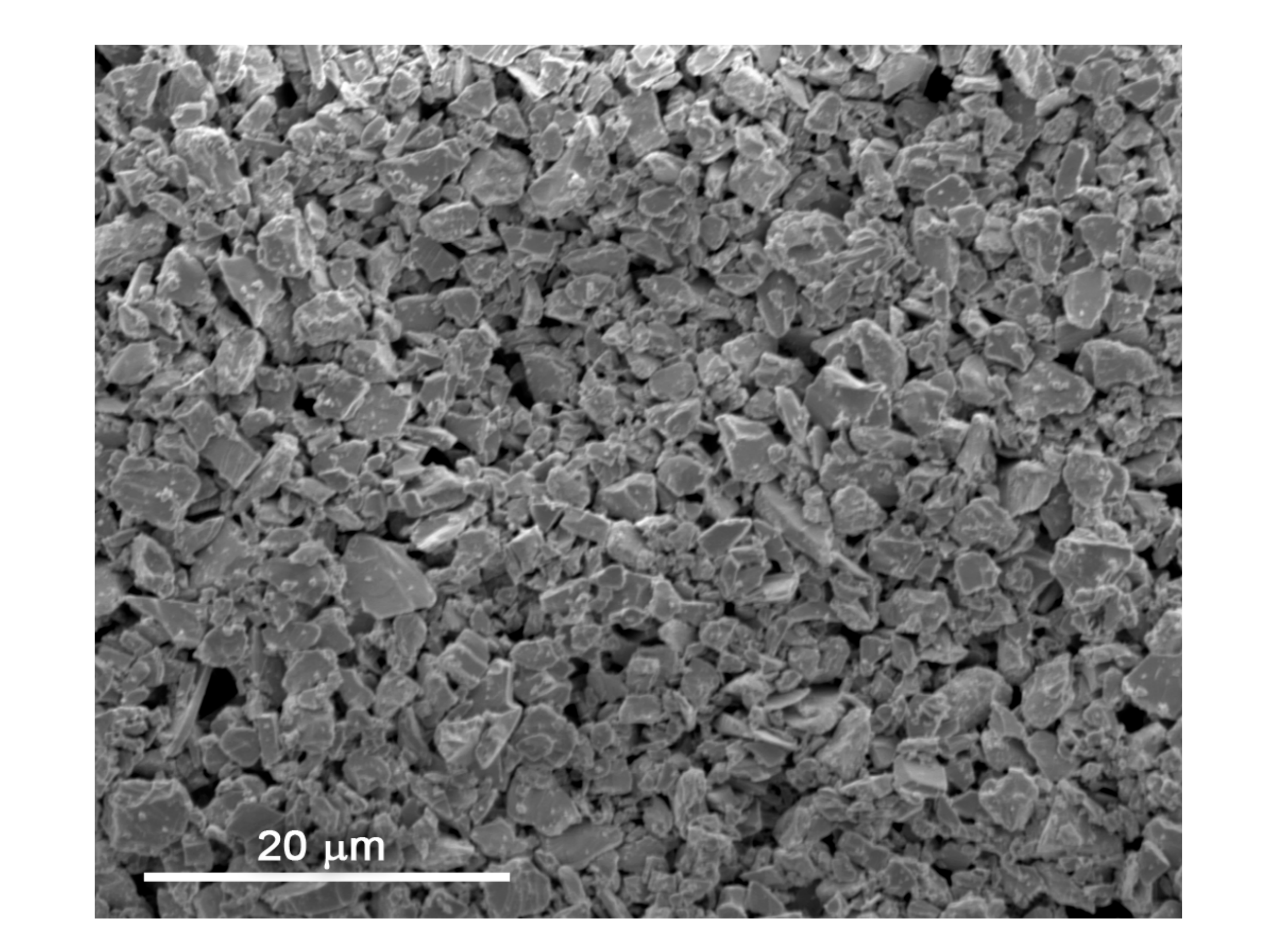}{b}
\includegraphics[width=7.5cm,angle=0]{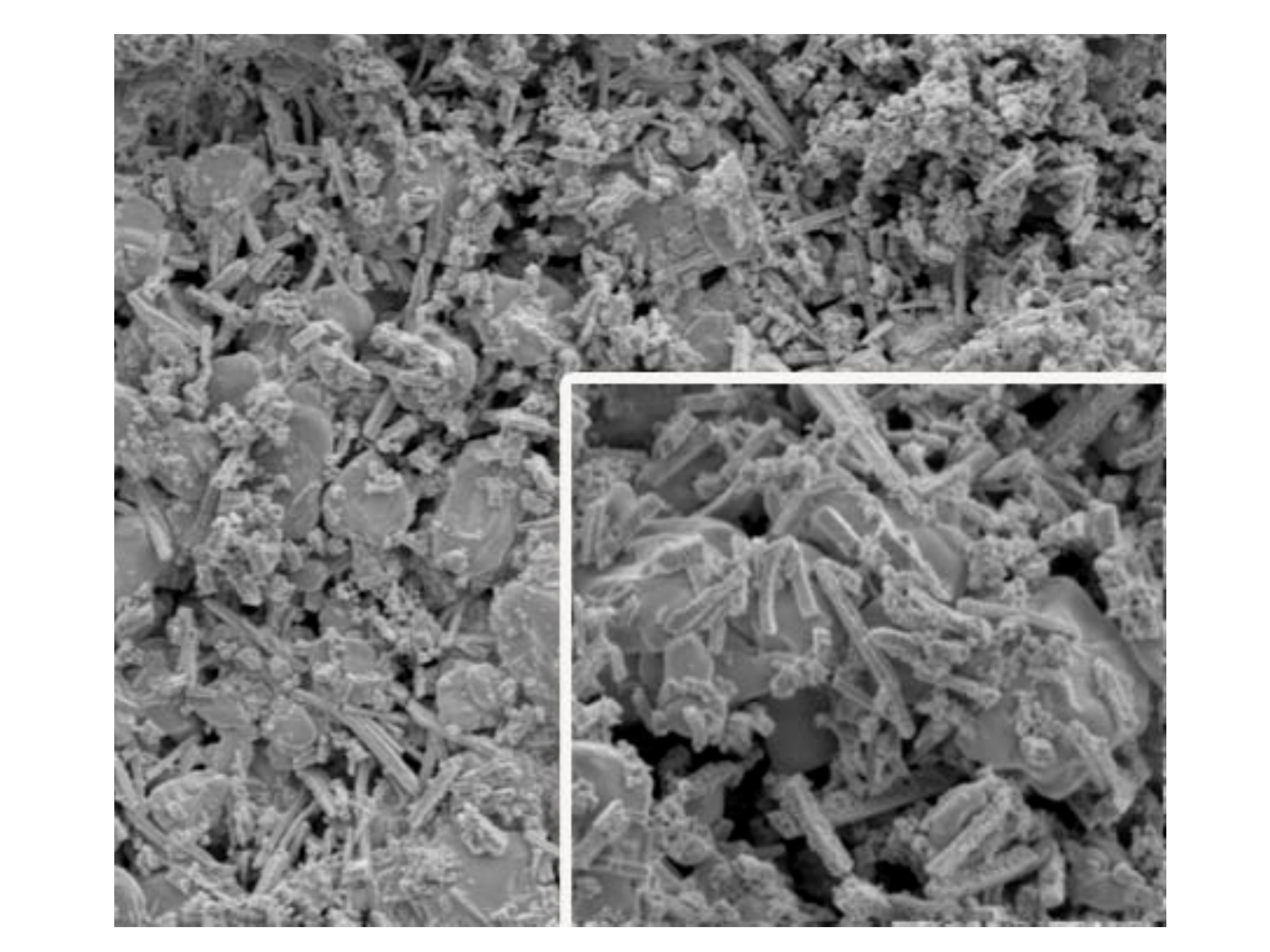}{c}
\caption{a) FE-SEM image of Mn$_{12}$-acetate ribbons. b) FE-SEM image of YBaCuO particles. c) FE-SEM image of the mixture of Mn$_{12}$-acetate ribbons and YBaCuO particles.}
\label{fig samples}
\end{figure}

Ribbon-shaped Mn$_{12}$-acetate particles were prepared by re-precipitation \cite{Imaz-ChemCom2008} of Mn$_{12}$-acetate crystals of size $4.9 \pm 1.0 \mu$m that were synthesized as previously described by Lis \cite{Lis}.  We first dissolved $60$-mg Mn$_{12}$-acetate crystals in $15$mL of acetonitrile, filtered the solution to avoid any solid trace, and mixed it with $30$mL of toluene under continuous stirring.  After one hour a brown solid precipitate was collected by filtration. Its field-emission scanning electron microscopy (FE-SEM) revealed the formation of ribbon-shaped particles, Fig. \ref{fig samples}-a. The average ribbon size was calculated statistically from FE-SEM images, measuring the length and the width of $150$ particles of the same sample. The calculated average length was $3.2 \pm 1.9 \mu$m with the median of $2.8 \mu$m, and the average width was $0.8 \pm 0.3 \mu$m with the median of $0.8 \mu$m. 

\begin{figure}[htbp!]
\includegraphics[width=9.5cm,angle=0]{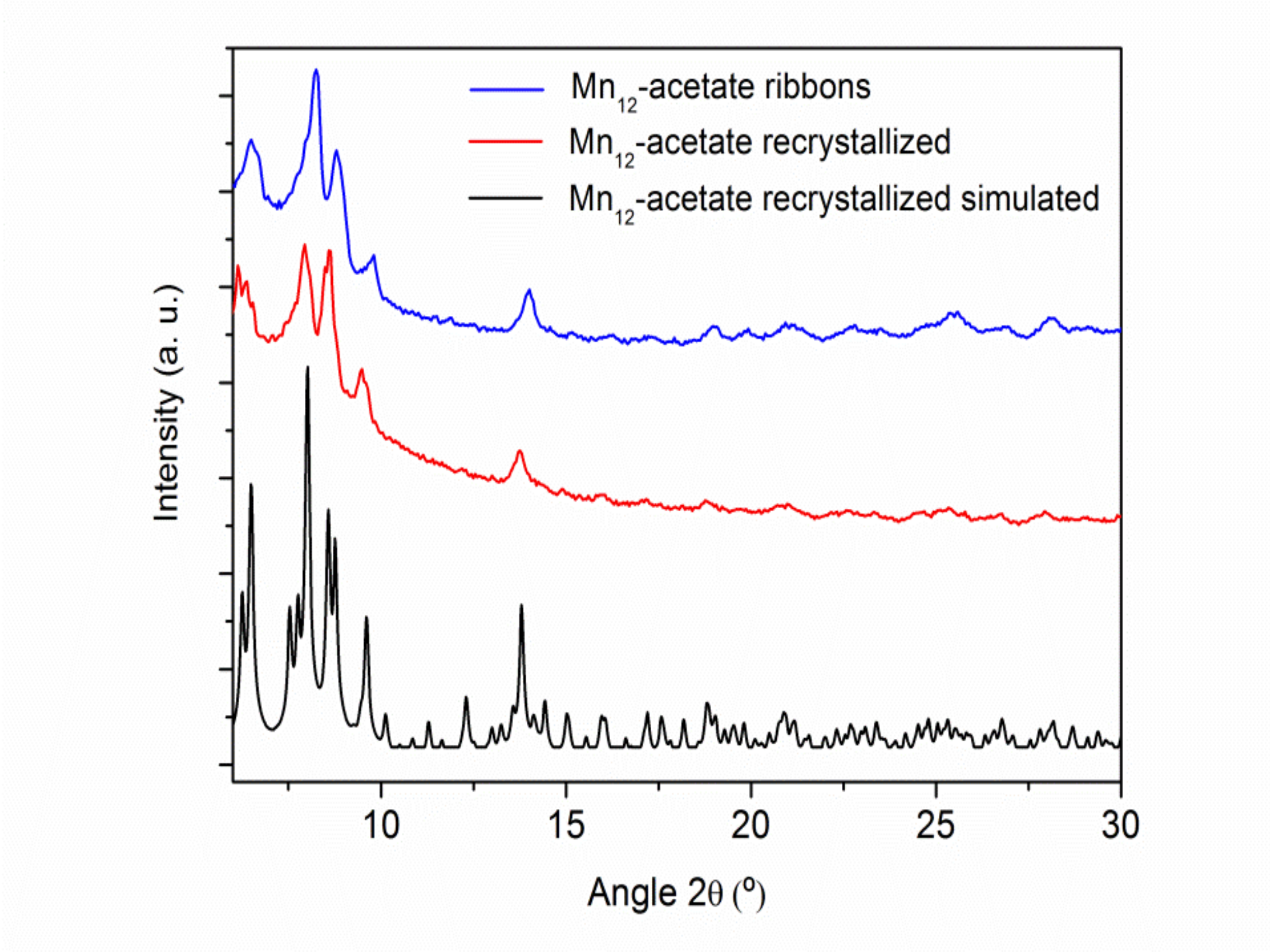}
\caption{Color online: X-ray data from M$_{12}$ ribbons (blue), from triclinic crystal grown as described in the text (red), and simulated X-ray pattern from a triclinic crystal (black).}
\label{fig X-ray}
\end{figure}

The X-Ray powder diffraction pattern of the ribbons showed packing of Mn$_{12}$ molecules that was different from the initial Mn$_{12}$-Ac crystals,  Fig. \ref{fig X-ray}. In order to grow single crystals of sufficient size to perform the single-crystal X-ray diffraction experiment the synthesis procedure was further modified. The initial Mn$_{12}$-acetate crystals were dissolved in acetonitrile, filtered, and subjected to slow diffusion of toluene vapors. After a few days, we observed the formation of rectangular shaped crystals. Due to their small size the single crystal X-ray diffraction experiment was performed under Synchrotron radiation in the XALOC beamline at the ALBA synchrotron. The powder pattern simulated from the resolution of the crystalline structure was in line with the powder diffractogram obtained from the ribbons, Fig \ref{fig X-ray}. The intramolecular structure was identical to that in a conventional Mn$_{12}$-Ac crystal of tetragonal symmetry. However, the intermolecular packing corresponded to the triclinic space group. 

Synthesis of the mixtures began with the preparation of a suspension of Mn$_{12}$-acetate particles in $3$mL of  toluene and a suspension of YBaCuO particles in the same amount of toluene. The size of YBaCuO particles, $2.9 \pm 1.3 \mu$m, was  calculated statistically from FE-SEM images, see Fig. \ref{fig samples}-b. Mixtures with different weight ratio were prepared by mixing the two suspensions under continuous stirring for 15 minutes. The resulting mixtures were filtered and the collected granular solids were ground for $10$ minutes to achieve an intimate mixing of Mn$_{12}$-acetate ribbons with YBaCuO particles. Field-Emission Scanning Electron Microscopy (FE-SEM) images demonstrated coexistence of both type of particles in the resulting mixtures, see Fig. \ref{fig samples}-c. 

\begin{figure}[htbp!]
\includegraphics[width=8.3cm,angle=0]{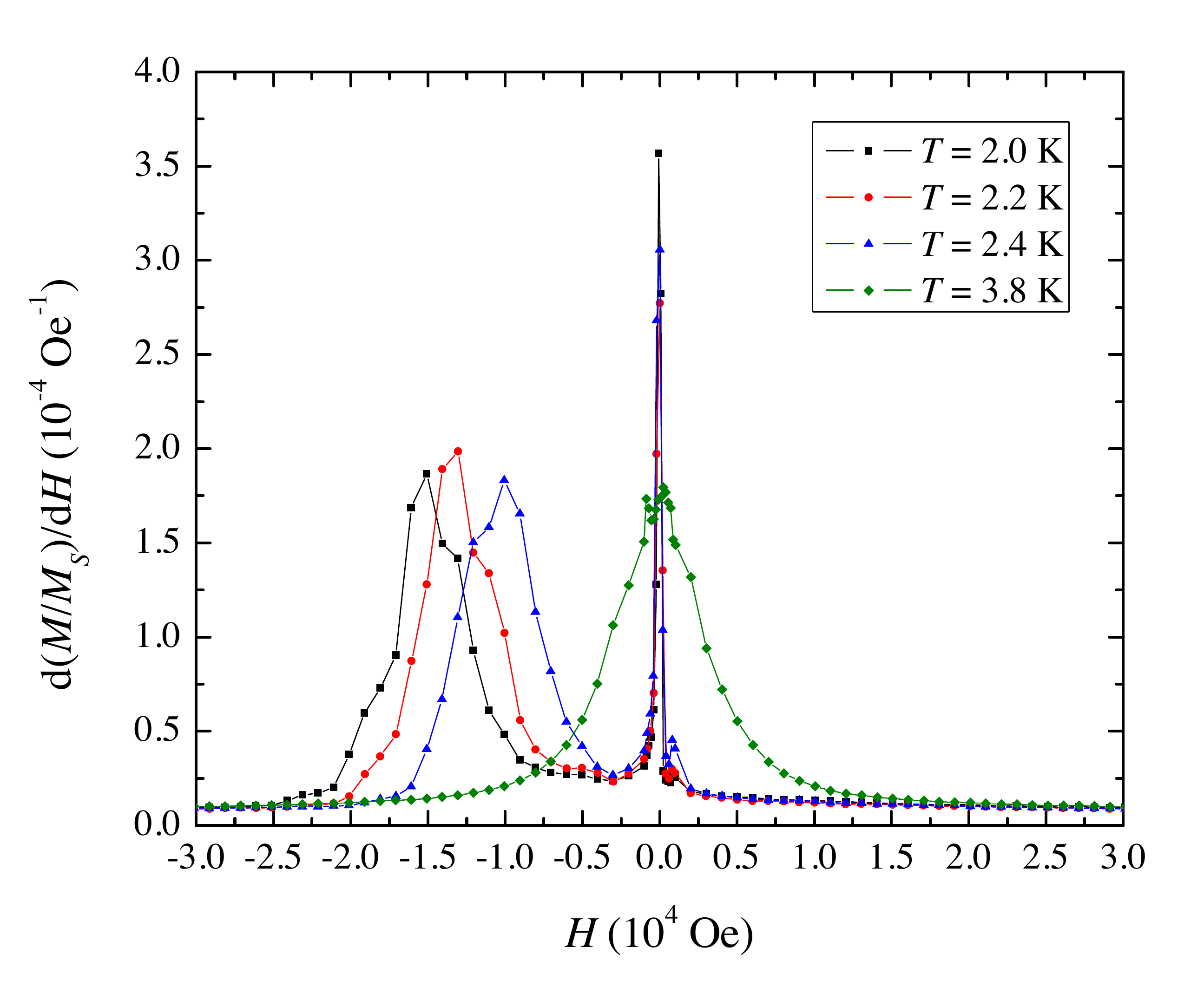}{a}
\includegraphics[width=8.3cm,angle=0]{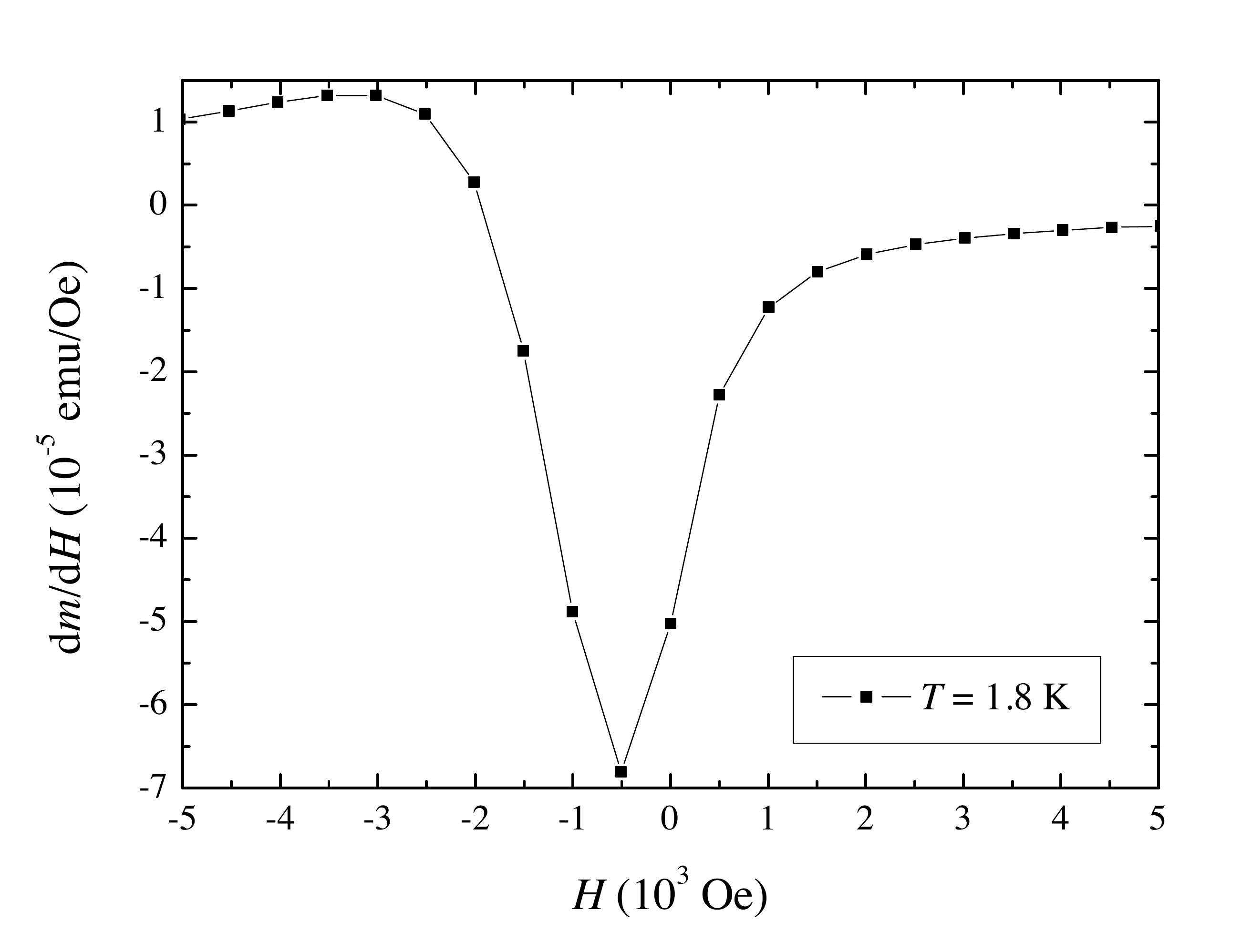}{b}
\caption{a) Field derivative of magnetization curves of the sample consisting of pure Mn$_{12}$ ribbons at different temperatures. Presence of conventional spin-tunneling maxima separated by $0.46$ T is apparent. b) Field derivative of the magnetization curve of the sample consisting of pure YBaCuO grains.}
\label{fig pure}
\end{figure}

Earlier we demonstrated \cite{Lendinez-PRB2015} that neither single crystals nor oriented microscrystals were needed to observe resonant spin-tunneling in molecular magnets. The tunneling maxima can be convincingly detected by plotting the field derivative of the magnetization curve measured in non-oriented or even amorphous microcrystals if the structure of the magnetic core of the Mn$_{12}$ molecules remains robust with respect to the local arrangement of molecules, which is the case for the ribbons. Fig. \ref{fig pure} shows field derivative of magnetization curves of Mn$_{12}$-acetate ribbons taken at different temperatures, and the field derivative of the magnetization curve of the sample consisting of pure YBaCuO grains. They clearly demonstrate that we deal with conventional Mn$_{12}$ molecules and conventional YBaCuO superconductor. 

\begin{figure}[htbp!]
\includegraphics[width=8.3cm,angle=0]{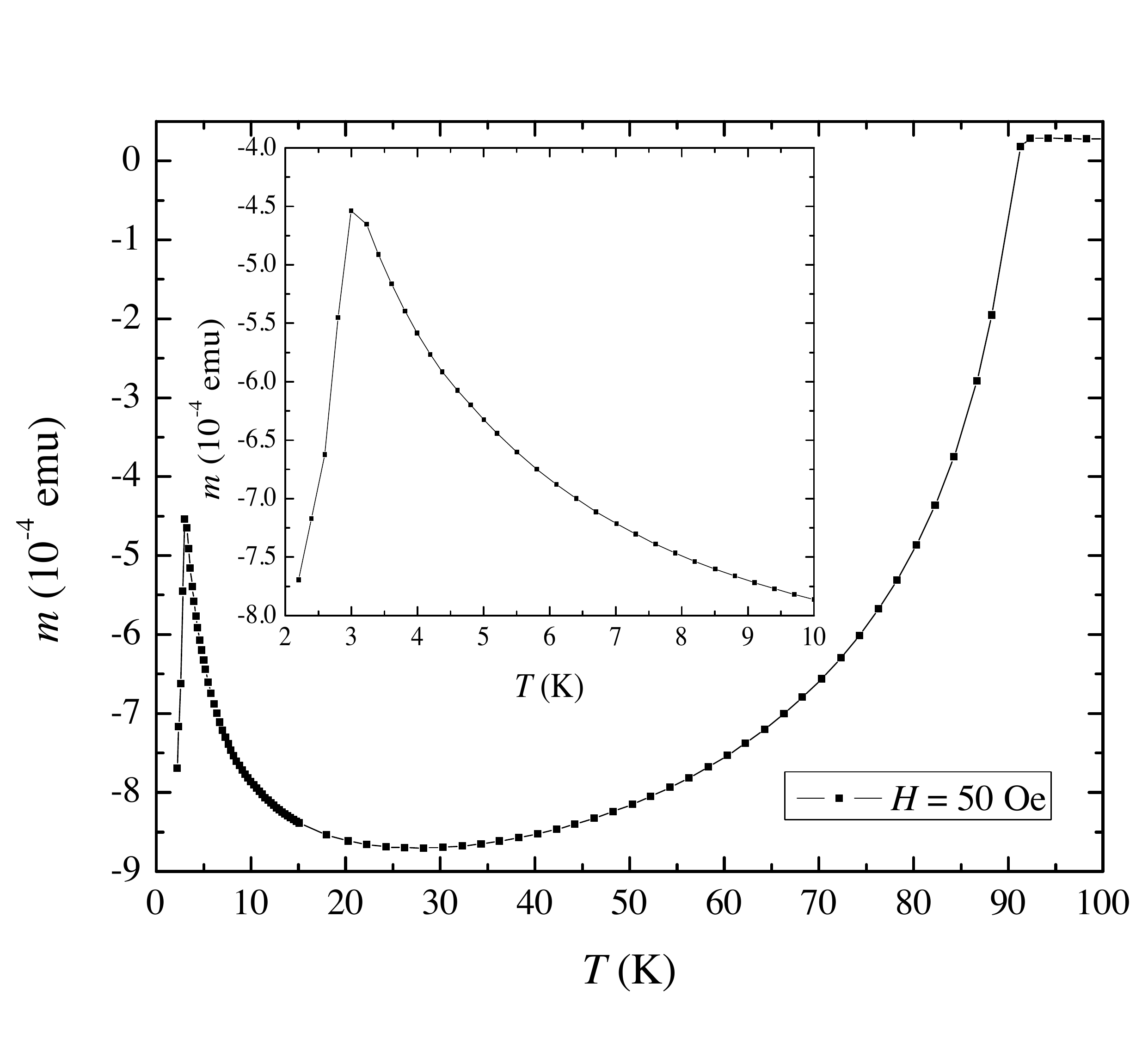}{a}
\includegraphics[width=8.3cm,angle=0]{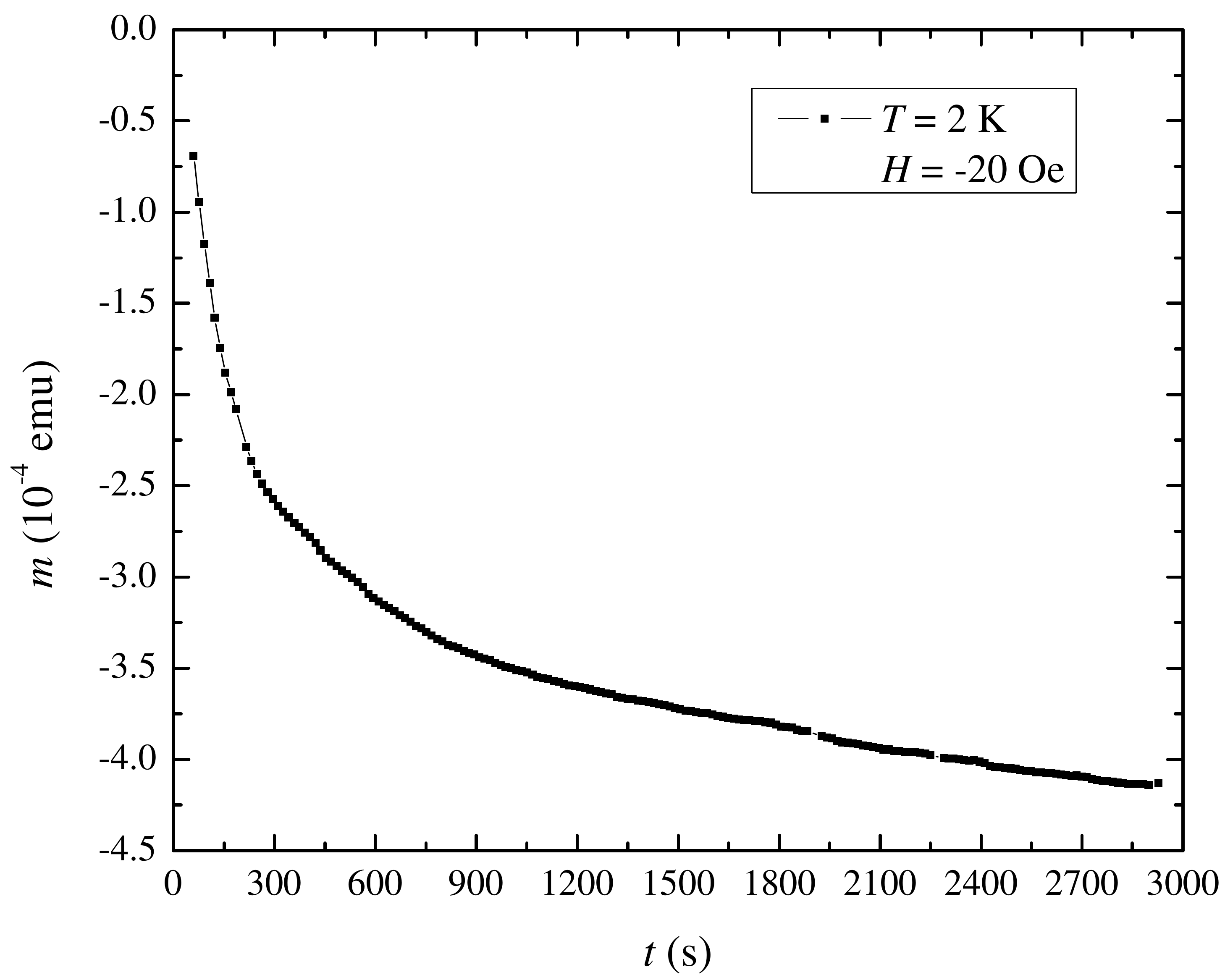}{b}
\caption{a) Zero-field-cooled magnetization curve of the 1:1 (equal masses) mixture of Mn$_{12}$-Ac ribbons and YBaCuO particles at $50$ Oe. The low-temperature paramagnetic peek (shown in the inset) is due to the ribbons. It  is superimposed on the diamagnetic signal from the superconductor. b) Magnetic relaxation from saturation to $-20$ Oe at $2.0$K.}
\label{fig ZFC-relaxation}
\end{figure}

Zero-field-cooled magnetization of curve of the 1:1 (equal mass) mixture of Mn$_{12}$-acetate ribbons and YBaCuO particles is presented in Fig. \ref{fig ZFC-relaxation}-a. It shows a pronounced conventional Mn$_{12}$ blocking maximum at $3.5$K and the absence of any second species of molecules with a different spin or a different magnetic anisotropy barrier. The paramagnetic moment of the ribbons is superimposed on the negative diamagnetic moment of YBaCuO, which makes the total moment negative. Fig. \ref{fig ZFC-relaxation}-b shows a typical magnetic relaxation of the 1:1 mixture at low field. The sample was initially magnetized in a $3$-T field at $T = 2$K. The field was first taken down to $1$ kOe and then rapidly reduced to zero and switched to $-20$ Oe, at which time the magnetization measurements were taken. The relaxation curve indicates an unusual for Mn$_{12}$ acetate, very rapid, large decrease of the magnetization on the time scale of a few minutes. 

\begin{figure}[htbp!]
\includegraphics[width=8.3cm,angle=0]{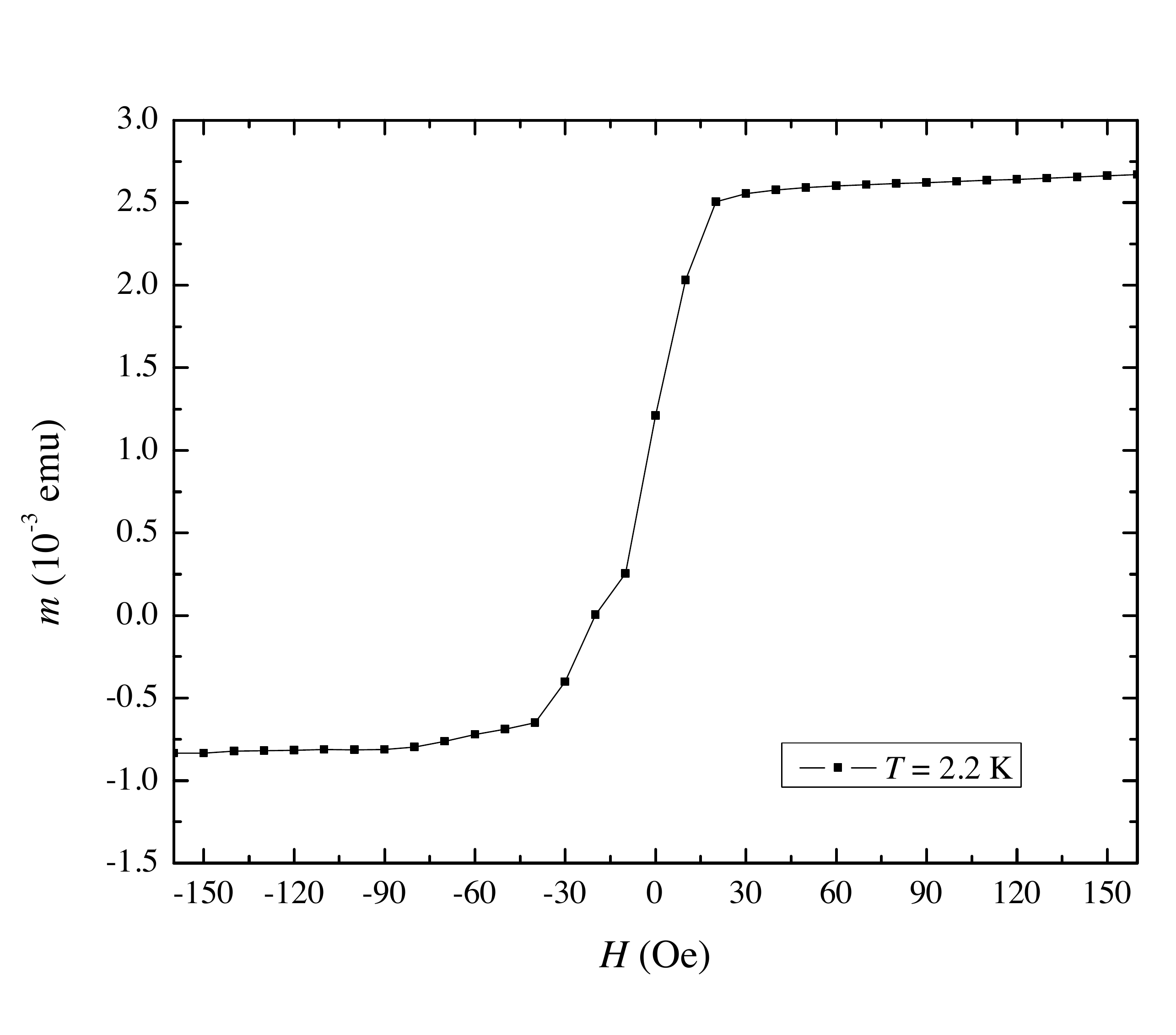}{a}
\includegraphics[width=8.3cm,angle=0]{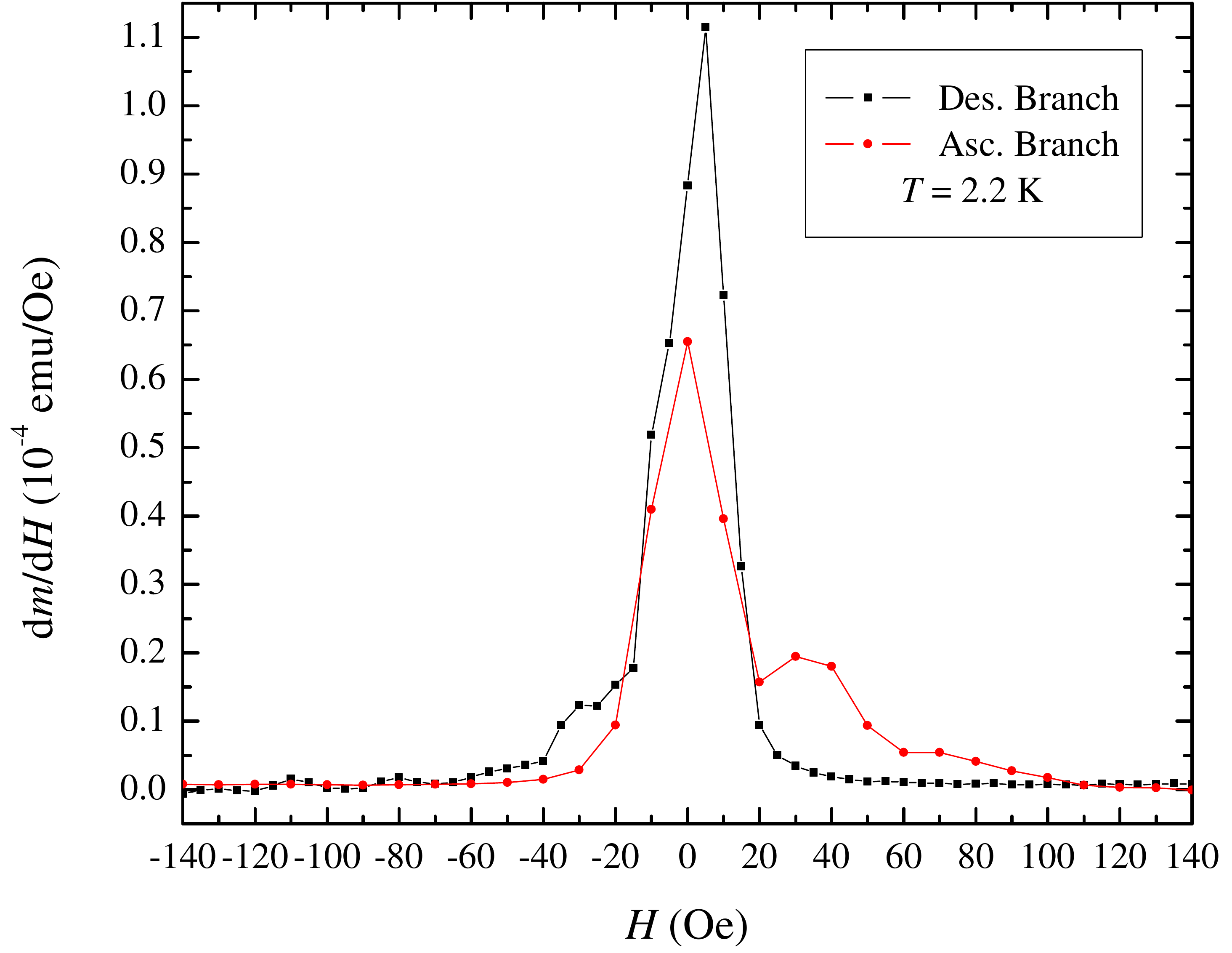}{b}
\caption{a) Low field part of the descending branch of the magnetization curve of the 1:1 (equal mass) mixture of Mn$_{12}$-Ac ribbons and YBaCuO particles at $T = 2.2$K. b) Field derivative of the descending (upper maximum) and ascending (lower maximum) of the magnetization curve.}
\label{fig magnetization}
\end{figure}

Low-field part of the descending branch of the magnetization curve of the $1:1$ mixture is shown in Fig. \ref{fig magnetization}-a. It represents an unusually large jump from the positive to negative magnetization in a narrow field interval from $20$ Oe to $-20$ Oe. Such jumps have only been seen in molecular magnets under the conditions of magnetic deflagration -- propagation of the front of combustion of the Zeeman energy which is similar to the combustion of a chemical substance \cite{CCNY-PRL2005,UB-PRL2005}.  However, in a zero field the Zeeman energy is zero and deflagration is not possible. Consequently, one has to assume that the jump is due to the unusually strong thermally-assisted quantum spin tunneling between matching spin levels. The derivative of the magnetization curve depicted in Fig. \ref{fig magnetization}-b shows that the width of the jump at half-height is in the ball park of $25$ Oe, which is smaller than the width of the zero-field resonance in the conventional pure Mn$_{12}$ acetate by more than one order of magnitude. The two maxima in Fig. \ref{fig magnetization}-b represent field derivatives of the descending (higher maximum) and ascending (lower maximum) of the magnetization curves. The ascending maximum is lower because a fewer number of molecules hold metastable direction of their magnetic moment after going through the descending branch. Note that in the field region of the maxima the magnetization of YBaCuO has a negative minimum (see Fig. \ref{fig pure}-b), which means that the maxima due to the resonant spin tunneling in the Mn$_{12}$ ribbons are even more dramatic than they appear. 

We should emphasize that the above data represent a small cut of the data accumulated by us on mixtures of different composition, taken at different temperatures, under different field-sweep protocols. The conclusion about the narrow width of the zero-field spin-tunneling resonance is consistent throughout all data.  We have checked that the observed magnetic relaxation comes entirely from the Mn$_{12}$ ribbons and not from the flux (if any) trapped in the superconducting grains. The sample prepared of YBaCuO grains alone showed no detectable relaxation at any temperature used in experiment. 

The only reasonable explanation to these findings is the screening of the dipolar fields in Mn$_{12}$ ribbons by superconducting currents. The irregular shape of the YBaCuO grains creates weak Josephson links between them in a compressed solidified mixture. At large field the links are broken by the field and the magnetization is a sum of the paramagnetic signal from the Mn$_{12}$ ribbons and diamagnetic signal from decoupled superconducting YBaCuO grains. As the field goes down and reaches a few tens of Oe the weak links come into play. Josephson currents begin to flow between the grains, screening the dipolar fields generated by the magnetic moments of the ribbons. As a result the dipolar broadening of the zero-field spin-tunneling resonance becomes suppressed. Resonances between spin energy levels with magnetic quantum numbers $\pm m$ are restored and thermally assisted quantum tunneling becomes greatly enhanced. 

In conclusion, we have demonstrated that mixing of a molecular magnet with a superconductor results in a significant narrowing of spin tunneling resonances due to the screening of dipolar fields by superconducting currents.

The work at the University of Barcelona has been supported by the Spanish Government Project No.  MAT2011-23698. A.G.-S. acknowledges support from Universitat de Barcelona. I.I. and J.E. thank the MINECO for the Ram\'{o}n y Cajal contract and the FPI fellowship, respectively. The work of EMC at CUNY Lehman College has been supported by the Office of Science of the U.S. Department of Energy through grant No. DE-FG02-93ER45487.

\end{document}